\documentclass[10pt,a4paper,twoside]{article}

\usepackage{times}
\usepackage[utf8]{inputenc}
\usepackage[T1]{fontenc}
\usepackage{graphicx}
\usepackage{url}
\usepackage{amsmath}
\usepackage{amsfonts}

\usepackage{taln2014}
\usepackage[frenchb]{babel}
\title{Algorithmes de classification et d'optimisation\,:\\
participation du LIA/ADOC à DEFT'14}

\author{Luis Adrián Cabrera-Diego\up{1,3}\quad Stéphane Huet\up{1} \quad Bassam Jabaian\up{1} \quad Alejandro Molina\up{1}\\ 
\quad Juan-Manuel Torres-Moreno\up{1,2} \quad Marc El-Bèze\up{1} \quad Barthélémy Durette\up{3}\\
  (1) LIA, UAPV, 91022 Chemin de Meinajariès, 84022 Avignon Cedex 9 \\ 
  (2) École Polytechnique de Montréal, Montréal, Québec Canada\\ 
  (3) ADOC Talent Management, 21 rue du Faubourg Saint-Antoine, 75011 Paris\\ 
cabrera@adoc-tm.com, 
\{bassam.jabaian,stephane.huet,juan-manuel.torres,marc.elbeze\}@univ-avignon.fr, alejandro.molina-villegas@alumni.univ-avignon.fr, durette@adoc-tm.com \\ 
}

\begin{document}

\maketitle

\resume{
L'édition 2014 du Défi Fouille de Textes (DEFT) s'est intéressée, entre autres, à la tâche d'identifier dans quelle session chaque article des conférences TALN précédents a été présenté.
Nous décrivons les trois systèmes conçus au LIA/ADOC pour DEFT 2014.
Malgré la difficulté de la tâche à laquelle nous avons participé, des résultats intéressants (micro-précision de 0,76 mesurée sur le corpus de test) ont été obtenus par la fusion de nos systèmes.
}
\\

\abstract{
This year, the DEFT campaign (Défi Fouilles de Textes) incorporates a task which aims at identifying the session in which articles of previous TALN conferences were presented. We describe the three statistical systems developed at LIA/ADOC for this task. A fusion of these systems enables us to obtain interesting results (micro-precision score of 0.76 measured on the test corpus).
}
\\

\motsClefs{classification de textes, optimisation, similarité}
{text classification, optimization, similarity}

\section{Introduction}

Dans le cadre de la conférence TALN 2014\footnote{\url{http://www.taln2014.org/site/}},
sera organisé en juillet 2014 à Marseille (Fran\-ce) un atelier centré sur un défi qui avait pour objet la fouille de textes. 
Ce défi est la dixième édition de DEFT (DÉfi Fouille de Textes). 
De notre côté, il s'agit de la sixième participation dans DEFT du Laboratoire Informatique d'Avignon (LIA)\footnote{\url{http://lia.univ-avignon.fr}}.
Cette fois-ci, le LIA a participé coinjointement avec l'entreprise ADOC Talent Management\footnote{\url{http://www.adoc-tm.com/}}.

La tâche 4 de DEFT a eté définie comme suit\footnote{\url{http://deft.limsi.fr/2014/index.php?id=2}}~:

\textsl{[...Cette tâche...] se démarque des précédentes car elle concerne les articles scientifiques présentés lors des dernières conférences TALN. Le corpus se composera des articles présentés en communication orale (ni poster, ni conférence invitée). Pour chaque édition, seront fournis~: un ensemble d'articles (titre, résumé, mots-clés, texte), la liste des sessions scientifiques de cette édition, et la correspondance article/session (sauf pour le test). Le corpus de test se composera d'une édition complète de TALN (articles et liste des sessions) pour laquelle il faudra identifier dans quelle session chaque article a été présenté.
}

L'objectif est donc de déterminer la session scientifique dans laquelle un article de conférence a été présenté.

\section{Prétraitement et normalisation du corpus d'apprentissage}

Le corpus d'apprentissage pour la tâche 4 est constitué par l'ensemble des articles scientifiques étiquetés par leur session et regroupé par année de publication. Pour chaque année, un fichier précise également le nombre d'articles présentés par session. Les articles scientifiques à traiter sont fournis dans des fichiers *.txt. Ils résultent d'une extraction du texte à partir du code source des fichiers PDF avec l'outil {\sl pdftotext}. Or, cette méthode a parfois l'inconvénient de générer différents types d'erreurs.

Un des problèmes les plus récurrents est celui du codage des lettres accentuées, le tréma de «\,ï\,» devenant par exemple «\,ı\"\,». On rencontre aussi certains problèmes au niveau de la préservation de la structure du texte. En effet, une phrase peut être découpée en plusieurs lignes ou une ligne peut contenir plusieurs phrases. De la même manière, des erreurs se produisent au niveau du découpage en paragraphes. Pour corriger ces erreurs nous utilisons les méthodes proposées par \cite{cabrerasegcv}. Les anomalies au niveau des accents sont repérées puis corrigées à l'aide d'expressions régulières. En ce qui concerne la structure, nous tenons compte de la ponctuation, des majuscules et des traits d'union afin de reconstituer des phrases et des paragraphes.

D'autres manipulations s'avèrent nécessaires pour obtenir de meilleurs résultats. 
L'élimination de symboles qui ne contiennent pas d'information sémantique, comme les lettres grecques ($\Sigma$, $\Pi$, $\Delta$), les noms des variables ($\lambda$, $x$, $t$) ou les caractères de contrôle du document (tabulation verticale, retour chariot...). 
Nous avons aussi uniformisé les différents caractères de citations (guillemets) à un seul type et les différents caractères d'union (traits d'union).

Nous avons réalisé une analyse automatique pour identifier les différentes parties de l'article\,: titre, auteurs, résumé, mots-clés, corps et références. Pour mener à bien cette tâche nous avons utilisé une méthode similaire à celle employée dans \cite{cabrerasegcv}, qui consiste à utiliser des expressions régulières pour trouver les différentes sections.

Lorsque l'article était rédigé en anglais, nous avons utilisé Google Traduction\footnote{\url{translate.google.com}.} pour les traduire automatiquement en français.

Après tous ces prétraitements, un seul fichier XML est produit avec la structure suivante\footnote{Ce fichier peut être consulté à l'adresse\,: \url{http://molina.talne.eu/deft2014training.xml}}~:
\begin{verbatim}
<?xml version="1.0" encoding="UTF-8" standalone="yes"?>
<deftcorpus type="training" year="2014">
<articles-2002>
  <article numArticle="taln-2002-long-007" normalisedSession="syntaxe">
    <title><![CDATA[...]]></title>
    <authors><![CDATA[...]]></authors>
    <abstract><![CDATA[...]]></abstract>
    <keywords><![CDATA[...]]></keywords>
    <body><![CDATA[...]]>
    </body>
    <biblio><![CDATA[]]>
    </biblio>
  </article>.
  ...
</articles-2002>
...
</deftcorpus>
\end{verbatim}

\section{Systèmes du LIA}

Nous avons développé trois systèmes indépendants que nous avons fusionnés par la suite\,:
\begin{enumerate}
\item Système collégial\,;
\item Système SimiPro\,;
\item Système à base de Champs Aléatoires Markoviens (CRF).
\end{enumerate}

\subsection{Système collégial}

Le premier système du LIA résulte de la fusion des avis émis
par 5 juges «\,indépendants\,» d'où le nom de système collégial.
Les approches qui leur sont attachées ont déjà été employées
par le LIA dans diverses campagnes d'évaluation comme DEFT \cite{el2007duel,torres2007,bost2013systemes},
ou REPLAB \cite{cossu2013lia} et ont été décrites dans les articles qui expliquent les modalités de notre participation à ces campagnes antérieures.
Parmi les 5 juges réunis pour participer à DEFT 2014, on trouve
une approche de type Cosine, un modèle probabiliste de type $n$-gramme
(avec $n$ variable), un modèle de Poisson, une similarité de type Jaccard
et enfin une approche de type k plus proche voisins.

Les paramètres de ces différents juges ont été entraînés en faisant
de la validation croisiée année par année sur le corpus d'apprentissage.
Ainsi le nombre de plus proches voisins a été fixé à 17 sur la globalité
du corpus. À l'issue des décisions prises par les 5 juges la méthode
de fusion qui a été employée est une fusion linéaire.

Enfin prenant appui sur les nombres de papiers propres à chaque
session, une réaffectation par permutations successives a été effectuée
pour maximiser le jugement moyen des 5 juges. Cette méthode ne garantit
pas l'obtention d'un optimum global.

\subsection{Système SimiPro}

Dans une conférence scientifique les articles acceptés sont regroupés en sessions. Les articles ayant une thématique similaire sont réunis dans une même session. 
Le LIA et ADOC Talent Management ont abordé la tâche 4 comme une tâche de similarité entre une représentation synthétique de chaque session et une représentation synthétique de chaque article. Dans la suite du propos, nous appellerons ces représentations respectivement «\,profil d'article\,» ($P_a$) et «\,profil de session\,» ($P_s$). 
Dans ce travail, un profil est l'ensemble de mots-clés qui représentent la thématique abordée. 
Les profils des sessions et des articles sont comparés en utilisant la distance cosinus. L'analyse de ces distances permet de classer les articles dans les sessions. Notre système est composé de quatre grandes étapes qui sont expliquées ci-après et sont appliquées sur le corpus de test et d'apprentissage.

La première étape est la lemmatisation et l'étiquetage morpho-syntaxique du corps des articles et des mots-clés indiqués par leurs auteurs. Son but est d'utiliser la forme canonique des mots et leur catégorie grammaticale dans les étapes suivantes. Cette étape est réalisée à l'aide du logiciel Freeling 3.1 \cite{padro12} pour le français.

La deuxième étape consiste à rechercher des séquences de mots ($c = m_1~m_2~\ldots~m_j$) susceptibles de constituer des mots-clés, c'est-à-dire des n-grammes de mots qui par leurs caractéristiques peuvent représenter les principales thématiques. Notre système utilise les catégories grammaticales identifiées par Freeling pour réaliser un découpage du texte. Ce découpage est réalisé au niveau des mots qui, de par leur catégorie grammaticale, sont peu susceptibles de constituer des mot-clés. Plus spécifiquement, le découpage est réalisé au niveau des verbes, des adverbes, des noms propres, des quantités, des conjonctions, des interjections, des dates et de la ponctuation \footnote{Freeling a la classe «\,F\,» pour représenter les symboles utilisés dans la ponctuation.}. La méthode appliquée ici est similaire à celle utilisée par le système d'extraction terminologique LEXTER \cite{bourigault1994lexter}. Pour réduire encore le nombre de mots-clés candidats, nous découpons également le texte au niveau des mots vides. Ce découpage a comme exception l'article «\,le\,» et la préposition «\,de\,» quand ils ne se trouvent pas aux extrémités d'une séquence de mots\footnote{Ces exceptions permettent au système de générer des séquences de mots comme «\,extraction de information\,» et «\,traitement de le parole\,».}. Un dernier filtre supprime les séquences de plus de 4 mots.

La troisième étape de SimiPro est le classement par pertinence des séquences de mots d'un article trouvées dans l'étape antérieure. Nous utilisons pour cela une version modifiée de l'algorithme \textit{Rapid Automatic Keyword Extraction} que nous avons réalisée pour DEFT 2014 de manière individuelle sur le corps de chaque article. La version originale de cet algorithme, appelé également RAKE, est décrite dans \cite{rose2010automatic}. L'algorithme RAKE calcule le degré de co-occurrence de chaque mot et leur nombre d'occurrences. Puis, il donne un poids $\mathbb{P}(m)$ à chaque mot $m$ en utilisant la formule suivante\,: 
\begin{equation}
  \mathbb{P}\left(m\right) = \left\{
	\begin{array}{l l l}
	  \frac{deg\left(m\right)}{f\left(m\right)} & \quad \text{si} & f(m) > 1\\
	  \frac{deg\left(m\right)}{F} & \quad \text{si} & f(m) = 1
	\end{array}\right.
\end{equation}
où $deg(m)$ est le degré de co-occurence de $m$, $f(m)$ le nombre d'occurrences de $m$ dans le texte et $F$ le nombre de mots dans le texte. Ensuite, pour chaque séquence de mots, $c = m_1~m_2~\ldots~m_j$, dans le texte, RAKE attribue le score $\mathbb{S}(c)$ suivant\,:
\begin{equation}
\mathbb{S}(c) = \frac{\sum_{i = 1}^{j}{\mathbb{P}(m_i)}}{j}\enspace .
\end{equation}
Notre version modifiée de RAKE donne des scores plus homogènes pour les séquences les plus petites\footnote{Dans la version originale les séquences les plus grandes ont par construction les scores les plus hauts.} ainsi que celles contenant des mots qui n'apparaîssent qu'une seule fois\footnote{De même, les séquences contenant des mots qui n'apparaîssent qu'une seule fois ont, par construction, un score élevé dans RAKE.}.

La quatrième étape de SimiPro est la création des profils. Cette étape varie selon que l’on considère la phase d’apprentissage ou la phase de test. Pendant la phase d’apprentissage, le système génère des profils de session. Pour créer un profil de session, SimiPro considère tous les mots-clés des articles affectés à une même session, indépendamment de l’année, pour générer une liste de mots-clés $\mathbb{L}$. Le système somme les scores des mots-clés apparaissant plusieurs fois. Le système crée autant de listes $\mathbb{L}$ que de sessions. Pendant la phase de test, SimiPro génère les profils des articles par année. Dans ce cas, le système débute en considérant l'ensemble de mots-clés de chaque article comme une liste $\mathbb{L}$.

Pour chacun des mots-clés dans les listes $\mathbb{L}$ générées précédemment selon la phase, le système applique une mesure basée sur le TF-IDF. La formule est la suivante :
\begin{equation}
\mathbb{V}(c) = \mathbb{S}(c) * log \left ( \frac{N}{n} \right )
\end{equation}
où $c$ est la séquence de mots, $\mathbb{V}$ le nouveau score de $c$, $\mathbb{S}$ le score de $c$ donné par RAKE, $N$ le nombre de documents dans la session dans la phase d'apprentissage, ou dans l'année dans la phase de test. La variable $n$ est le nombre de documents de la session ou de l'année où $c$ apparaît selon le cas.

Pour filtrer plus facilement les valeurs de $\mathbb{V}$ avec un seuil entre 0 et 1, nous avons décidé d'appliquer deux types de normalisations. Le système commence par une normalisation cosinus \cite{salton1988term}. Chaque liste $\mathbb{L}$ est considérée comme un vecteur $\bar{l}$ dans le modèle vectoriel. Pour chaque vecteur $\bar{l}$, le système obtient leur norme $||\bar{l}||$. Puis, SimiPro calcule leur vecteur unitaire $\bar{l}_u = \bar{l}/||\bar{l}||$. Ce vecteur unitaire a comme composantes les scores normalisés $\mathbb{V}_n$. La deuxième normalisation divise les valeurs de chaque composantes de $\bar{l}_u$ par la valeur maximale de toutes les composants du vecteur afin d'obtenir un score,$\mathbb{V}_{n2}$, compris 0 et 1 et ayant une même échelle pour tous les mots-clés de toutes les listes. À l'issue de cette phase de normalisation, les mots-clés des auteurs sont ajoutés aux listes normalisées avec un score de 1.

Le profil d'une session, $P_s$, ou d'un article $P_a$, selon le cas, est généré à partir de leur liste respective de mots-clés normalisée. Seules les séquences de mots qui ont une valeur $\mathbb{V}_{n2}$ supérieure ou égale à 0,50 sont prises en compte.

Pendant la phase de test et une fois les profils des articles créés, SimiPro débute la classification des articles parmi les sessions. Le système charge, premièrement, les profils $P_a$ des articles à classer d'une année déterminée. Puis, il charge les profils $P_s$ des sessions correspondant à cette même année\footnote{Pour chaque année dans le corpus de test les organisateurs ont fournis le nom des sessions et le nombre d'articles par session.}. Dans le cas d'une nouvelle session n'ayant pas d'équivalent dans le corpus d'apprentissage, le système crée un profil en utilisant l'algorithme suivant : 1/ Le nom de la session $s$ est lemmatisé avec Freeling et est considéré comme un mot-clé d'un nouveau profil du type $P_s$. Si $s$ est multi-mots, on prend également le premier et le dernier mot (par exemple : détection de thèmes $\rightarrow{}$ détection, thèmes ; taln pour le tic $\rightarrow{}$ taln, tic). 2/ On sélectionne les profils de session contenant $s$. 3/ Le système cherche $s$ dans tous les profils $P_s$ créés pendant l'apprentissage. Il génère la liste $A$ contenant les sessions où $s$ a été trouvée et la liste $B$ contenant les séquences de mots contenant $s$. 4/ La liste $B$ de mots-clés est ajoutée au nouveau profil. 5/ On croise les mots clés de tous les profils de la liste A et on conserve tous les mots-clés apparaissant plus d'une fois.

Dès que SimiPro a chargé tous les profils, le système calcule la similarité entre les profils de sessions $P_s$ et les profils d'articles $P_a$. Ce calcul est réalisé en utilisant la mesure cosinus. Les articles sont finalement classés parmi les sessions par ordre décroissant de score de similarité en prenant en compte les contraintes fournies concernant le nombre d'articles par sessions.

\subsection{Système basé sur les CRF} 

Les CRF («\,Conditional Random Fields\,» ou «\,Champs Aléatoires Markoviens\,») sont une famille de modèles graphiques introduite par \cite{lafferty2001}. Ils ont le plus souvent été utilisés dans le domaine du traitement automatique des langues, pour étiqueter des séquences d'unités linguistiques.  Ces modèles possèdent à la fois les avantages des modèles génératifs et des modèles discriminants de l'état de l'art. En effet, comme les classifieurs discriminants, ils peuvent manipuler un grand nombre de descripteurs, et comme les modèles génératifs, ils intègrent des dépendances entre les étiquettes de sortie et prennent une décision globale sur la séquence.

La représentation de ces modèles est décrite comme une probabilité conditionnelle d'une séquence de concept $ C = c_{1}, ... , c_{T}  $ étant donnée une séquence de mots $ W = w_{1}, ... , w_{T} $. Cette probabilité peut être calculée comme suit :  
\[ P(C|W)= \frac 1Z  \prod_{t=1}^T  H(c_{t-1},c_{t},\phi(w_{1}^{N},n))\]
avec
\[ H(c_{t-1},c_{t},\phi(w_{1}^{N},n)) = \exp (\sum_{m=1}^{M} \lambda_{m} \cdot h_{m}(c_{t-1},c_{t},\phi(w_{1}^{N},n)))\]
$H$ est un modèle log-linéaire fondé sur des fonctions caractéristiques $ \ h_{m}(c_{t-1},c_{t},\phi(w_{1}^{N},n)) $ qui représentent l'information extraite du corpus d'apprentissage. Cela peut être par exemple $ \ w_{t-2}^{t+2}$ représentant une fenêtre de voisinage de taille 2 autour du mot courant. Les poids $ \lambda$ du modèle log-linéaire sont estimés lors de l'apprentissage et $ \ Z $ est un terme de normalisation appris au niveau de phrases. 

Afin de bénéficier des avantages de ces classifieurs discriminants, nous avons proposé de considérer la tâche 4 de DEFT comme un problème d'étiquetage d'un document source, sauf que les étiquettes possibles sont toutes les sessions de la conférence.  

L'apprentissage d'un modèle CRF nécessite des données annotées (étiquetées) au niveau des mots, ainsi chaque mot de chaque article représente une observation pour le modèle. Considérer la totalité de l'article pour la construction du modèle CRF peut être très coûteux en terme de complexité d'autant plus qu'un bruit important peut être intégré dans le modèle à cause du nombre important de mots communs entre différents articles de différentes sessions. Pour minimiser ce bruit, nous avons fait le choix de sélectionner l'information à prendre en compte dans le modèle et donc pour chaque article seules les données liées aux titre, résumé, noms d'auteurs et mots-clés sont prises en compte.

Pour l'apprentissage, nous considérons uniquement les articles appartenant à des sessions présentes parmi les sessions du test. Cette règle, qui a pour but d'empêcher l'étiquetage d'un article par une session non attendue dans le test, élimine une part très importante des données d'apprentissage.  Pour éviter cette élimination massive, une projection de sessions vers des sessions similaires a été réalisée. Par exemple la session «\,traduction|alignement\,» est remplacée par «\,alignement\,», «\,recherche d'information\,» est projetée  vers «\,extraction d'information\,».
  
L'outil Wapiti \cite{lavergne2010} a été utilisé pour apprendre les modèles CRF en considérant des fonctions (features) uni-grammes avec une fenêtre de voisinage de taille 2 autour du mot courant et des fonctions bi-grammes. L'algorithme de descente de gradient a été appliqué pour optimiser les poids du modèle.  

Une fonction uni-gramme permet de prendre en compte une seule étiquette à la fois caractérisant l'association du mot et de l'étiquette. Les fonctions bi-grammes portent sur un couple d'étiquettes successives. Ce type de fonction permet par exemple d'apprendre des informations sur une liste de mots-clés ou sur les coauteurs d'une session donnée.  

Pour le test, nous avons considéré le même type d'information que pour l'apprentissage du modèle (titre, résumé, auteurs et mots-clés). Lors du décodage, le modèle attribue à chaque mot une étiquette avec un score de probabilité. Pour un article donné, la somme des log-probabilités des mots étiquetés par la même session a été calculée et l'article est affecté à la session qui correspond à la somme maximale de ces log-probabilités.

\subsection{Fusion et optimisation}

Les trois systèmes précédents fournissent un score $s_{ij}^{(k)}$ d'association d'un article $i$ à une session $j$. De manière à faciliter leur combinaison, les scores de chaque système $k$ sont normalisés entre 0 et 1. Les résultats produits par les trois systèmes sont combinés à l'aide d'une interpolation linéaire\,:
\begin{equation}
s_{ij} = \sum_{k=1}^3 \lambda_k s_{ij}^{(k)}\enspace .
\end{equation}
Les poids $\lambda_k$ sont optimisés à l'aide d'une recherche sur grille avec des pas de 0,05 pour maximiser la macro-précision sur un corpus développement, en retenant pour chaque article $i$ la session $j$ qui obtient le meilleur score $s_{ij}$.

La tâche~4 de DEFT, pour laquelle sont fournis le nombre d'articles $n_j$ par session $j$, peut être vue comme un problème d'optimisation linéaire discrète (OLD) consistant à trouver les valeurs $x_ij$ satisfaisant la fonction objectif~(\ref{eq:ILPobj}).
\begin{eqnarray}
 \max \sum_{i=1}^m \sum_{j=1}^n x_{ij} s_{ij} \label{eq:ILPobj}\\
 \sum_{j=1}^n x_{ij} = 1 & i=1...m \label{eq:ILPart}\\
 \sum_{i=1}^m x_{ij} = n_j & j=1...n \label{eq:ILPses}\\
 x_{ij} \in \{0,1\} & i=1...m, j=1...n \nonumber
 \end{eqnarray}
Les contraintes (\ref{eq:ILPart}) s'assurent que chaque article $i$ n'est associé qu'à une seule session, tandis que les égalités (\ref{eq:ILPses}) représentent les contraintes quant au nombre d'articles par session. Nous avons employé cette modélisation notée OLD en post-traitement de chaque système ou de leur combinaison par interpolation linéaire. Par rapport à la méthode classique ne retenant que la session $j$ qui maximise un score $s_{ij}$ pour chaque article $i$ indépendamment des autres articles, l'optimisation linéaire permet de prendre une décision au niveau d'une année. Le problème OLD est résolu à l'aide de l'algorithme du simplexe.

\section{Expériences sur le développement}

Les articles fournis pour les années 2008 et 2011 ont été utilisés comme corpus de développement pour régler les différentes paramètres de nos systèmes et les poids $\lambda_i$ du modèle d'interpolation linéaire. Le tableau~\ref{tab:devRes} reporte les résultats obtenus sur 2008 et 2011.

\begin{table}[th]
   \centering
   \begin{tabular}{l|cc|cc}
	\hline
    & \multicolumn{2}{c|}{2008} & \multicolumn{2}{c}{2011}\\
	& & +OLD & & +OLD\\
    \hline
Collégial   & 0,72 & 0,79 & 0,39 & 0,47\\
SimiPro     & 0,63 & 0,69 & 0,40 & 0,30\\
CRF         & 0,39 & 0,44 & 0,40 & 0,50\\
Fusion      & 0,62 & 0,77 & 0,46 & 0,37\\
    & \multicolumn{2}{c|}{$\lambda_k$ : 0,4 ; 0,05 ; 0,55} & \multicolumn{2}{c}{$\lambda_k$\,: 0,05 ; 0,95 ; 0,0}\\
    \hline
   \end{tabular}
   \caption{\label{tab:devRes} Macro-précision mesurée sur les articles de 2008 et 2011 en optimisant respectivement les poids de l'interpolation linéaire sur 2011 et 2008.}
\end{table}

La comparaison des colonnes 2 et 3 d'une part et 4 et 5 d'autre part montrent que la prise en compte des contraintes sur le nombre d'article grâce à l'optimisation linéaire discrère améliorent les résultats pour tous les systèmes à l'exception de SimiPro et de la fusion pour l'année 2011.

Les poids utilisés pour combiner les systèmes sur 2008 et 2011 sont respectivement indiquées aux colonnes 2 et 3 de la dernière ligne du tableau~\ref{tab:devRes} et montrent une grande variabilité entre ces deux corpus. Cette instabilité entre les deux années se reflète dans les valeurs de macro-précision obtenu par la combinaison (dernière ligne) qui ne sont supérieures au meilleur système que pour l'année 2011 sans le post-traitement OLD.

Les poids finalement retenus pour le corpus de test ($\lambda_k$ : 0,1 ; 0,8 ; 0,1) sont optimisés sur l'ensemble des articles de 2008 et 2011 pour augmenter la généralisation du modèle de combinaison.

\section{Résultats}

Sur la tâche 4  il y a eu 5 participants pour un total de 13 soumissions. 
Les systèmes ont été évalués en utilisant les mesures proposées par DEFT.
Les cinq meilleures soumissions varient (la meilleure de chaque équipe) de 0,2778 à 1,000 (mesure\,: précision à 1). Moyenne=0,5926\,; Médiane=0,4815\,; Écart-type=0,2860.

Le tableau~\ref{tab:testRes} présente nos résultats obtenus sur le corpus de test. Nous constatons que l'optimisation linéaire discrète permet des gains, notamment lors de la fusion de nos trois systèmes (+0,19 points au niveau de la macro-précision).

\begin{table}[th]
   \centering
   \begin{tabular}{l|cc}
	\hline
	& & +OLD \\
    \hline
Collégial    & 0,57 & 0,66\\
SimiPro      & 0,45 & 0,34\\
CRF          & 0,50 & 0,51\\
\bf Fusion       & 0,54 & \bf 0,73\\
	\hline
   \end{tabular}
   \caption{\label{tab:testRes} Macro-précision (au niveau des sessions) mesurée sur le corpus de test.}
\end{table}

Pour la campagne, trois systèmes ont été soumis : 
le résultat de la fusion avec optimisation linéaire discrète (Tableau~\ref{tab:resSoumission}, ligne~1), le système SimiPro avec optimisation linéaire discrète (ligne~2) et le système collégial avec une optimisation locale par permutation (ligne~3). Comme attendu, la fusion conduit aux meilleurs résultats, même si ceux-ci restent très proches du système collégial.

\begin{table}[th]
   \centering
   \begin{tabular}{l|ccc}
    \hline
              & 2012 & 2013 & 2012 \& 2013\\
    \hline
\bf Fusion + OLD  &\bf 0,77 (0,76) &\bf 0,75 (0,70) &\bf 0,76 (0,76)\\
SimiPro + OLD & 0,32 (0,32) & 0,41 (0,36) & 0,37 (0,36)\\
Collégial + optimisation locale & 0,68 (0,67) & 0,72 (0,65) & 0,70 (0,70)\\
	\hline
   \end{tabular}
   \caption{\label{tab:resSoumission} Micro-précision (macro-précision au niveau des années) mesurées sur le corpus de test pour nos trois soumission pour la campagne.}
\end{table}

\section{Conclusions et travail futur}

Malgré les différences entre le corpus d'apprentissage et le test, notamment au niveau du nom des sessions, nos algorithmes ont conduit à des résultats très intéressants, bien au dessus de la moyenne des participants. 
La fusion s'est averée une stratégie performante, car elle a su combiner les résultats des 3 systèmes, en surpassant ceux du meilleur d'entre eux.
Par manque de temps, nous n'avons pas intégré d'autres composants TAL dans nos approches (entités nommées, resumé automatique, etc). Nous pensons qu'un système de résumé automatique guidé \cite{favre2006lia,torres2011resume} pourrait être utilisé dans ce cadre de manière à mieux cibler les passages contenant les mots-clés. 
Également, nous considérons que l'extraction de mots-clés pourrait être effectuée en utilisant des algoritmes performants basés sur les graphes \cite{mihalcea2004textrank}.

\section*{Remerciements}

Nous voulons remercier l'ANRT par le financement de la convetion CIFRE n$^\circ$~2012/0293b entre ADOC Talent Management et l'Université d'Avignon et des Pays de Vaucluse, ainsi que le \textit{Consejo Nacional de Ciencia y Tecnología (CONACyT)} du Mexique (bourse n$^\circ$ 327165).

\bibliographystyle{taln2002}
\bibliography{biblio}

\end{document}